\documentclass[twocolumn,english]{revtex4-1}
\usepackage[T1]{fontenc}
\usepackage[latin9]{inputenc}
\usepackage{array}
\usepackage{multirow}
\usepackage{amsmath}
\usepackage{amssymb}
\usepackage{graphicx}
\usepackage{esint}

\makeatletter

\providecommand{\tabularnewline}{\\}

\makeatother

\usepackage{babel}
\begin{document}

\title{Maximizing the hyperpolarizability of 1D potentials with multiple
electrons}

\author{Christopher J. Burke}

\affiliation{Department of Physics and Astronomy, Tufts University, 574 Boston
Avenue, Medford, Massachusetts 02155, USA}

\author{Joseph Lesnefsky}

\affiliation{Department of Physics, Case Western Reserve University, 10900 Euclid
Avenue, Cleveland, Ohio 44106, USA}

\author{Rolfe G. Petschek}

\affiliation{Department of Physics, Case Western Reserve University, 10900 Euclid
Avenue, Cleveland, Ohio 44106, USA}

\author{Timothy J. Atherton}

\email{timothy.atherton@tufts.edu}

\affiliation{Department of Physics and Astronomy, Tufts University, 574 Boston
Avenue, Medford, Massachusetts 02155, USA}
\begin{abstract}
We optimize the first and second intrinsic hyperpolarizabilities for
a 1D piecewise linear potential dressed with Dirac delta functions
for $N$ non-interacting electrons. The optimized values fall rapidly
for $N>1$, but approach constant values of $\beta_{int}=0.40$, $\gamma_{int}^{+}=0.16$
and $\gamma_{int}^{-}=-0.061$ above $N\gtrsim8$. These apparent
bounds are achieved with only 2 parameters with more general potentials
achieving no better value. In contrast to previous studies, analysis
of the hessian matrices of $\beta_{int}$ and $\gamma_{int}$ taken
with respect to these parameters shows that the eigenvectors are well
aligned with the basis vectors of the parameter space, indicating
that the parametrization was well-chosen. The physical significance
of the important parameters is also discussed. 
\end{abstract}
\maketitle

\section{Introduction}

Nonlinear optical materials are the active constituent for many applications
such as light modulators, contrast agents for medical imaging and
therapy, optical solitons, phase conjugation mirrors and optical self-modulation.
In each of these, the performance of the system is improved by using
a material with a stronger nonlinear response, quantified by various
nonlinear susceptibilities defined by expanding the induced polarization
$P$ in a power series in the applied electric field,
\begin{equation}
P=\alpha E+\beta EE+\gamma EEE+O(\epsilon^{4}).
\end{equation}
Here, $\alpha$ is the linear susceptibility familiar from dielectric
materials; $\beta$ and $\gamma$ are the nonlinear susceptibilities
and are referred to as the first and second hyperpolarizability respectively.
These quantities are in general frequency-dependent tensors that depend
on the electronic structure of the constituent molecules, their symmetry,
ordering and the material in which they are embedded. In the present
work we focus on the off-resonant molecular contribution. Much effort
has been expended over the years in synthesizing new molecules with
higher $\beta$ or $\gamma$. Comparisons between materials must be
made carefully however, because these quantities increase trivially
with the size of the molecule.

Important progress on developing suitable figures-of-merit for comparison
was made by Kuzyk, who showed that fundamental quantum mechanics requires
that $\beta$ and $\gamma$ for the off-resonant case are bounded
by the inequalities,
\begin{eqnarray}
 & \left|\beta\right| & \le\sqrt[3]{4}\left(\frac{e\hbar}{\sqrt{m}}\right)^{3}\frac{N^{3/2}}{E_{10}^{7/2}}\equiv\beta_{0}^{max},\\
\left(\frac{e\hbar}{\sqrt{m}}\right)^{4}\frac{N^{2}}{E_{10}^{5}}\le & \gamma & \le4\left(\frac{e\hbar}{\sqrt{m}}\right)^{4}\frac{N^{2}}{E_{10}^{5}}\equiv\gamma_{0}^{max}\label{eq:kuyzklimit}
\end{eqnarray}
where $E_{10}$ is the difference between the ground and first excited
state and $N$ is the number of electrons participating. From the
maximum values $\beta_{0}^{max}$ and $\gamma_{0}^{max}$, one defines
intrinsic hyperpolarizabilities,
\begin{equation}
\beta_{int}=\beta/\beta_{0}^{max},\ \gamma_{int}=\gamma/\gamma_{0}^{max}.
\end{equation}
The intrinsic quantities have the property that they remain invariant
under a simultaneous rescaling of energy and length,
\begin{equation}
x\to x'E^{1/2},\ V(x)\to V'(x')E\label{eq:rescaling}
\end{equation}
and hence are useful quantities for comparing materials because they
remove the irrelevant scaling with size. Analysis of extant materials
following the derivation of the bounds in eq. (\ref{eq:kuyzklimit})
revealed that all of them fell short of the fundamental limits by
more than an order of magnitude, an observation that has catalyzed
a great deal of research over the past decade on how to create materials
that approach these fundamental limits. The derivation of the bounds
provides some guidance\textemdash for optimal $\beta$ and $\gamma$
only three states are assumed to significantly contribute to the hyperpolarizabilities
and the optimum can be achieved by tuning the dipole transition matrix
elements and energy level spacings\textemdash but does not construct
an explicit molecule or potential that has these properties. 

Subsequent work, thoroughly reviewed in \cite{kuzykreview} has attempted
to determine whether these predictions are universal and to explicitly
construct potentials that approach them. One approach has been to
conduct Monte Carlo searches of Hamiltonians with arbitrary spectra
and dipole transition elements to identify those with large $\beta_{int}$
and $\gamma_{int}$. The results support the three-state hypothesis,
though the optima found in such calculations need not correspond to
a local potential. To address this, several authors have numerically
optimized the intrinsic hyperpolarizabilities with respect to the
potential function for 1 electron, using various representations of
the potential, including power laws\cite{Mossman2013}, elementary
functions with a superimposed Fourier series\cite{Watkins2012}, piecewise
linear potentials\cite{Atherton2012,burke2013} and quantum graphs\cite{lytel2013a,lytel2013dressed,lytel2014optimum,lytel2015}.
The best potentials from these different studies possess hyperpolarizabilities
within the bounds of eq. (\ref{eq:kuyzklimit}) with the best known
values found of $\beta_{int}\sim0.71$ and $\gamma_{int}\sim0.60$
achieved in several studies with qualitatively different potentials.
It has therefore been speculated that the fundamental limits may require
exotic potentials and not be achievable with local potential functions.
The effect of including multiple electrons on the optimized potentials
has, however, received relatively little attention. Watkins and coworkers
\cite{watkins2011} found for $N=2$ electrons that the best intrinsic
hyperpolarizabilities are somewhat lower than for the $N=1$ electron
case, but even with electron-electron interactions included, the universal
features identified in other studies remained the same. 

In this paper, we apply the potential optimization technique to potentials
with $N>2$ electrons that interact only through Pauli exclusion.
This is a key step towards simulating realistic molecules. We find
that the best values of $\beta_{int}$ and $\gamma_{int}$ fall off
with increasing $N$ from the $N=1$ electron case, but rapidly converge
to a universal value. The small number of parameters in our potentials
allows a detailed exploration of the ``landscape'' of $\beta_{int}$
and $\gamma_{int}$ around the maximum. As in previous work, the hyperpolarizabilities
are more sensitive to one parameter than the other. Dimensional and
approximate analytical arguments allow us to provide physical interpretations
of these two parameters in terms of the wavefunctions of the highest
occupied molecular orbital. 

The paper is organized as follows: in section \ref{sec:Model} the
choice of potential, calculation and optimization techniques are briefly
reviewed; in section \ref{sec:Results} we present results for $\beta_{int}$
and $\gamma_{int}$ separately together with some discussion of the
implications of our results for identifying the features of potentials
most important to the hyperpolarizabilities; brief conclusions are
presented in section \ref{sec:Conclusion}.

\section{Model\label{sec:Model}}

\begin{figure}
\includegraphics{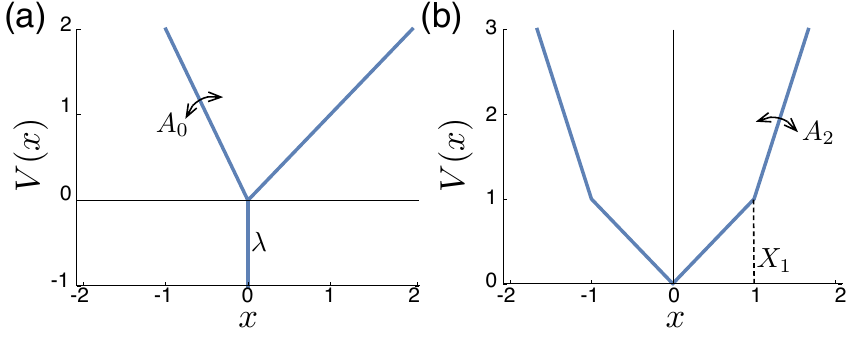}

\caption{\label{fig:Schematic}Schematics of the potential classes in which
$\beta_{\mathrm{int}}$ and $\gamma_{\mathrm{int}}$ are optimized.
(a) Asymmetric triangle will with a delta function at the center.
The slope on the right side is fixed at $1$ while the slope of the
side $A_{0}$ and the strength of the delta function $\alpha$ can
be varied. $A_{0}$ can be fixed at 1 to study a symmetric 1-parameter
potential. (b) a 2 parameter symmetric linear piecewise potential.
The first slope to the right of the origin is fixed at 1. The position
of the boundary between the first and second elements $X_{1}$ and
the second slope $A_{2}$ can be varied. The left side of the potential
is constrained to be the reflection of the right side.}
\end{figure}

Following a similar approach to that established in our earlier papers\cite{Atherton2012,burke2013},
we optimize $\beta_{int}$ and $\gamma_{int}$ with respect to the
shape of a piecewise linear potential dressed with Dirac delta functions
for $N$ electrons interacting only through Pauli exclusion. We perform
this optimization for two carefully chosen potentials depicted in
Fig. \ref{fig:Schematic} as well as a more general potential. The
first type is an asymmetric triangular well with a delta function
at the center {[}fig. \ref{fig:Schematic}(a){]}, 
\begin{equation}
V_{1}(x)=-\alpha\delta(x)+\begin{cases}
-A_{0}x, & x<0\\
x & x\ge0
\end{cases},\label{eq:potentialdelta}
\end{equation}
parametrized by the left hand slope $A_{0}>0$ and the strength of
the $\delta$-function $\alpha$. The effect of the $\delta$ function
is to introduce a sudden change in the phase of the wavefunction.
Lytel \emph{et al. }\cite{lytel2015} recently showed that the addition
of a $\delta$ function to a 1D potential has an equivalent effect
on the wavefunction to adding a side chain on a quantum graph. This
correspondence suggests that 1D potentials dressed with $\delta$-functions
could be engineered in molecules by the addition of appropriate side
groups. 

The second type of potential we consider, depicted in fig. \ref{fig:Schematic}(b),
contains only linear elements defined for $x>0$, 
\begin{equation}
V_{2}(x)=\begin{cases}
x & 0<x<X_{1}\\
A_{2}(x-X_{1})+X_{1} & x\ge X_{1}
\end{cases},\label{eq:potentiallinear}
\end{equation}
and for $x<0$ defined by enforcing $\mathcal{P}$ symmetry, i.e.
$V(-x)=V(x)$; this potential is specified by two parameters $X_{1}$
the position of the boundary between the two elements and $A_{2}$
the slope of the outermost element. These potentials were motivated
by our results in \cite{Atherton2012,burke2013} that only 2 parameters
at most were important to the optimization of both $\beta_{int}$
and $\gamma_{int}$; they have been designed to achieve the known
limits for $N=1$ electrons with residual flexibility. For example,
we showed in \cite{burke2013} that a triangular well with a $\delta$-function
of variable strength at the center, i.e. taking the potential (\ref{eq:potentialdelta})
and fixing $A_{0}=1$, was able to reach within $3\%$ of the upper
bound for $\gamma_{int}$ with $\lambda$ as the only free parameter.
The parametrization has also been chosen to eliminate variables irrelevant
to $\beta_{int}$ and $\gamma_{int}$ associated with translations
of the potential and rescalings of the form (\ref{eq:rescaling}). 

We also minimized $\beta_{int}$ for a piecewise linear potential
with $m$ elements,
\begin{equation}
V(x)=\begin{cases}
A_{0}x+B_{0} & x<x_{0}\\
A_{n}x+B_{n} & x_{n-1}<x<x_{n}\\
A_{m}x+B_{n} & x>x_{m-1}
\end{cases},\label{eq:GeneralPotential}
\end{equation}
with positions $x_{n}$ and slopes $A_{n}$ as the parameter set describing
the potential. We used this potential in our earlier paper on maximizing
$\beta_{int}$ for one electron\cite{Atherton2012}. There are some
necessary constraints on the parameters: the $x_{n}$ are strictly
ascending; $x_{0}=0$ and $B_{0}=0$ with no loss of generality and
$B_{1}=B_{0}$ with the remaining constants $B_{n}$ given by,
\begin{equation}
B_{n}=\sum_{m=1}^{n-1}(A_{m}-A_{m+1})x_{m}.
\end{equation}
The energy scale associated with the potential is also fixed, as in
the previous paper, by setting $A_{1}=1$. Having imposed these constraints,
there remain $2N-1$ free parameters. 

For each of these potentials, $\beta_{int}$ and $\gamma_{int}$ were
calculated for $N$ electrons as follows: first, the Schr\"odinger
equation is written for each segment as,
\begin{equation}
\left[-\frac{1}{2}\frac{d}{dx^{2}}+(A_{n}+\epsilon)x+B_{n}\right]\psi_{n}=E\psi_{n}
\end{equation}
where $A_{n}$ and $B_{n}$ are the slope and offset in the $n$th
segment and $\epsilon$ is the applied electric field. This can be
solved analytically using the well-known Airy functions,
\begin{eqnarray}
\psi_{n}(x) & = & C_{n}\text{Ai}\left[\frac{\sqrt[3]{2}\left(B_{n}-E+x\left(A_{n}+\epsilon\right)\right)}{\left(A_{n}+\epsilon\right)^{2/3}}\right]\nonumber \\
 &  & +D_{n}\text{Bi}\left[\frac{\sqrt[3]{2}\left(B_{n}-E+x\left(A_{n}+\epsilon\right)\right)}{\left(A_{n}+\epsilon\right)^{2/3}}\right].
\end{eqnarray}
To solve for the coefficients $C_{n}$ and $D_{n}$ in each element,
a set of boundary conditions are assembled at the edge of each element
from the usual conditions, i.e., 
\begin{eqnarray}
\psi_{n+1}(X_{n})-\psi_{n}(X_{n}) & = & 0\\
\psi'_{n+1}(X_{n})-\psi_{n}'(X_{n}) & = & \alpha_{n}\psi_{n}(X_{n})
\end{eqnarray}
where $\alpha_{n}$ is the strength of the delta function centered
at $X_{n}$. The requirement that $\psi\to0$ as $x\to\pm\infty$
eliminates two coefficients. The boundary conditions can be written
as a set of linear equations,
\begin{equation}
W\cdot u=0
\end{equation}
where $u$ is a vector comprised of the $C_{n}$ and $D_{n}$ coefficients
and $W$ is a matrix that depends on $E$, $\epsilon$ and the parameters
$A_{n}$ and $X_{n}$. The single electron energy levels $\lambda_{i}$
for the potential are determined by numerically finding the roots
of,
\begin{equation}
\det W=0,\label{eq:Determinant}
\end{equation}
setting $\epsilon=0$. Having determined these, we construct the non-interacting
$N$ electron ground state from the single electron states by successively
filling the energy levels using the \emph{aufbau} principle; we similarly
determine the first excited state by promoting an electron from the
highest occupied orbital to the lowest unoccupied orbital. In this
work we focus only on even values of $N$, as this simplifies determining
which electron to promote. The $N$-electron ground state energy
$E_{0}$, and that of the first excited state $E_{1}$, are then determined
by summing over the energies of the individual single electron energies,
\begin{equation}
E_{n}=\sum_{i}\nu{}_{i}^{n}\lambda_{i},\label{eq:EnergyOccupance}
\end{equation}
where $\nu_{i}^{n}$ is the occupation number for the $i$-th single
electron level in the $n$-th multi-electron state. 

The hyperpolarizabilities $\beta$ and $\gamma$ are obtained by differentiating
the ground state energy as a function of $\epsilon$,
\begin{equation}
\beta\equiv\frac{1}{2}\frac{d^{3}E_{0}}{d\epsilon^{3}},\ \gamma\equiv\frac{1}{6}\frac{d^{4}E_{0}}{d\epsilon^{4}}.
\end{equation}
The necessary derivatives can be related to the single-electron energy
levels by differentiating (\ref{eq:EnergyOccupance}) with respect
to $\epsilon$. An important advantage of the piecewise linear potential
is that the necessary derivatives of the single electron energy levels
$\lambda_{i}$ are conveniently obtained by repeated differentiation
of the determinant eq. (\ref{eq:Determinant}) using the Jacobi formula,
\[
\frac{d}{d\epsilon}\det W=\text{Tr}\left(\text{adj}W\cdot\frac{dW}{d\epsilon}\right),
\]
where $\text{adj}W$ is the adjugate matrix of $W$, together with
the chain rule, 
\begin{equation}
\frac{dW}{d\epsilon}=\frac{\partial W}{\partial\epsilon}+\frac{\partial W}{\partial\lambda_{i}}\frac{d\lambda_{i}}{d\epsilon}.
\end{equation}
To avoid repetition, formulae are available for these derivatives
in references \cite{Atherton2012} and \cite{burke2013}. Having evaluated
these derivatives, the intrinsic hyperpolarizabilities are easily
calculated numerically and a program to do so was implemented in \emph{Mathematica
10}. In subsequent sections, we present the optimized results as a
function of $N$, together with the corresponding best potentials.

\section{Results\label{sec:Results}}

\subsection{First hyperpolarizability}

\begin{figure}
\begin{centering}
\includegraphics[width=1\columnwidth]{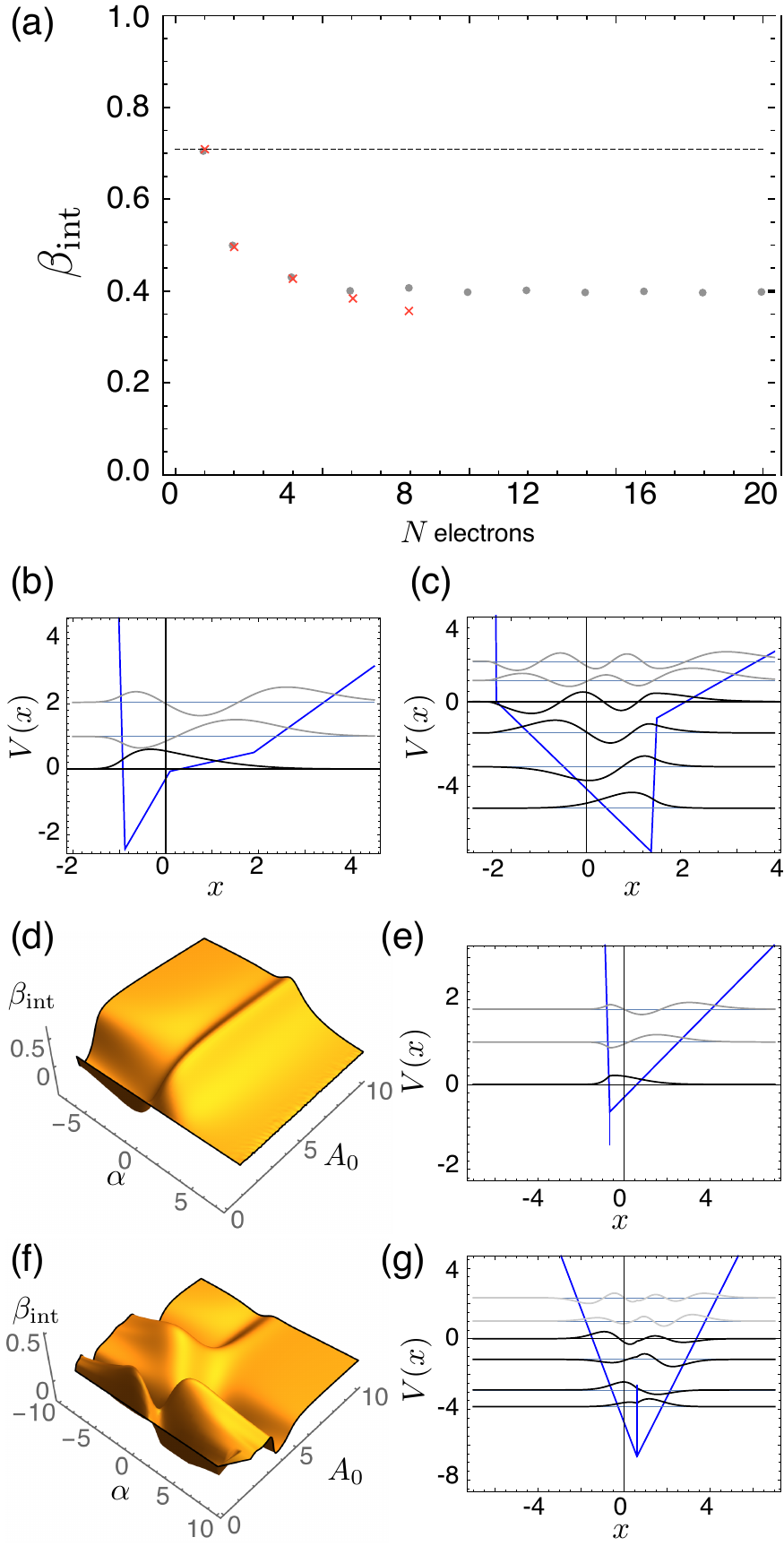}
\par\end{centering}

\caption{\label{fig:betaresults}(a) Maximum $\beta_{\mathrm{int}}$ achieved
in 2-parameter asymmetric $\delta$-function potential. Results for
the asymmetric delta function potential are shown as grey circles;
results for an arbitrary linear potential are shown as red crosses.
The dashed line represents the largest values of $\beta_{\mathrm{int}}$
for one electron found to date. Optimized potential and wavefunctions
for the arbitrary linear potential with (b) $N=2$ electrons and (c)
$N=8$ electrons. For the asymmetric $\delta$-function potential
and $N=2$, (d) the objective functions $\beta_{\mathrm{int}}$ plotted
versus the $\delta$-function potential shape parameters $A_{0}$
and $\alpha$ and (e) the optimized potential and wavefunctions. Corresponding
plots for $N=8$ electrons are shown in (f) and (g). For all subfigures,
wavefunctions plotted in black are occupied in the ground state; those
in grey correspond to the two lowest unoccupied states. }
\end{figure}

\begin{table*}
\begin{centering}
\begin{tabular}{>{\centering}m{0.5in}>{\centering}m{1in}>{\centering}m{0.5in}>{\centering}m{0.5in}>{\centering}m{0.75in}>{\centering}m{0.75in}>{\centering}m{1in}>{\centering}p{0.5in}>{\centering}p{0.5in}}
\hline 
Value of $\beta_{int}$ & Potential and wavefunctions & $A_{0}$ & $\alpha$ & Hessian Eigenvalues & Hessian Principal Eigenvector & $x_{nm}$ & $X$ & $E$\tabularnewline
\hline 
\hline 
0.404 (100\%) & \includegraphics[width=1in]{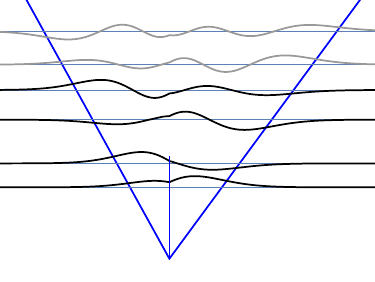} & 1.334 & -2.229 & -6.84, -0.099  & $\left(\begin{array}{c}
0.999\\
0.035
\end{array}\right)$ & \includegraphics[width=1in]{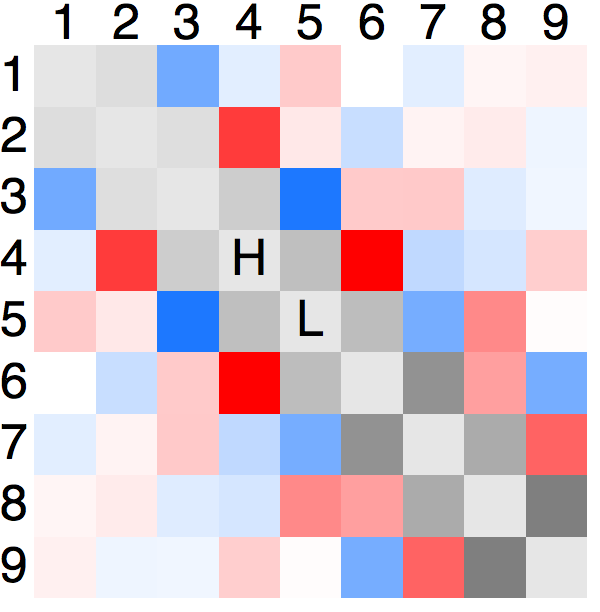} & 0.573 & 0.500\tabularnewline
-0.372 (96\%) & \includegraphics[width=1in]{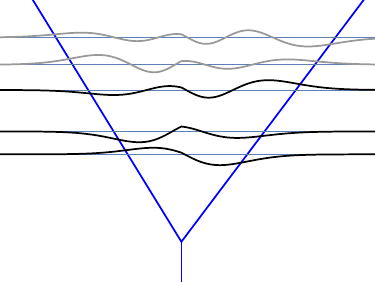} & 1.227 & 1.928 & 7.70, 0.130 &  $\left(\begin{array}{c}
0.999\\
0.027
\end{array}\right)$ & \includegraphics[width=1in]{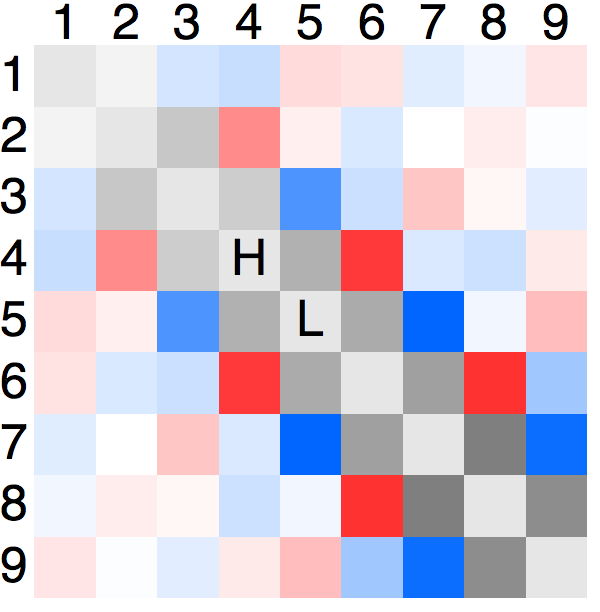} & 0.570 & 0.500\tabularnewline
0.370 (92\%) & \includegraphics[width=1in]{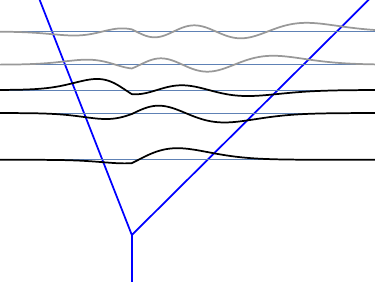} & 2.568 & 2.307 & -1.24, -0.014  & $\left(\begin{array}{c}
0.986\\
-0.162
\end{array}\right)$ & \includegraphics[width=1in]{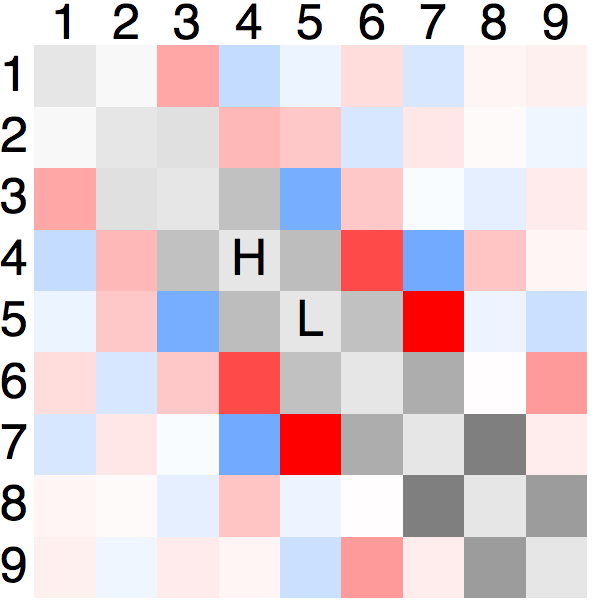} & 0.569 & 0.523\tabularnewline
\hline 
\end{tabular}
\par\end{centering}

\caption{\label{tab:Beta}Asymmetric $\delta$-function potentials with locally
optimal $\beta_{int}$. Results for each potential are shown columnwise:
the value of $\beta_{int}$ , and the fraction of the globally optimum
value; the potential and wavefunctions with black lines indicating
states that are occupied and light gray lines indicated unoccupied
states in the ground state configuration; optimized values of the
parameters $A_{0}$ and $\delta$; the eigenvalues and eigenvectors
of the Hessian matrix of $\beta_{int}$ with respect to the parameters;
a plot of the dipole transition matrix $x_{nm}=\left\langle n\left|x\right|m\right\rangle $
with the Highest Occupied Molecular Orbital (HOMO) and Lowest Unoccupied
Molecular Orbital (LUMO) indicated by H and L respectively, coloring
described in the text; the energy ratio $E$ and dipole transition
moment $X$ for the Kuzyk three-state model. }
\end{table*}

We optimized $\beta_{int}$ with respect to the parameters of the
asymmetric $\delta$-function potential of eq. (\ref{eq:potentialdelta})
as well as the more general piecewise linear potential with of eq.
(\ref{eq:GeneralPotential}) with four linear elements and $6$ parameters.
The number of linear elements was chosen because earlier work showed
no improvement in $\beta_{int}$ after 6 parameters\cite{Atherton2012}.
The highest values found for increasing $N$ are displayed in fig.
\ref{fig:betaresults}(a) for both of these. It is apparent that the
best achievable result diminishes for $N>2$ but rapidly reaches a
plateau of $\beta_{int}\sim0.4$ and that both potentials give consistent
results. 

For $N\le8$ electrons, the two choices of potential give very consistent
results, suggesting that we have indeed found a likely global optimum.
For $N>8$ electrons, however, the optimization procedure failed to
find a maximum of $\beta_{int}$ for the general linear potential
that approached that of the asymmetric $\delta$-function potential.
We speculate that this is because the number of local maxima increases
with $N$; we shall show that this is true later for the more carefully
chosen potentials at least. Because of this, we did not consider the
general piecewise linear potential further but display the optimized
potential and wavefunctions for $N=2$ and $N=8$ respectively in
fig. \ref{fig:betaresults}(b) and (c). That these potentials achieve
similar values of $\beta_{int}$ to the asymmetric $\delta$-function
potentials, despite visually appearing very different, supports our
conclusion in previous work\cite{Atherton2012} that $\beta_{int}$
is poorly determined in potential space with many irrelevant directions. 

Due to the small parameter space of the asymmetric $\delta$-function
potential, it is possible to directly visualize the objective function;
this is displayed in fig. \ref{fig:betaresults}(d) for $N=1$ or
2 electrons (the magnitude is scaled by $2^{-1/2}$ for the $N=2$
electron case), with corresponding optimized potential and wavefunctions
shown in fig. \ref{fig:betaresults}(e). The objective function for
$N=8$ electrons and the optimized potential and wavefunctions are
shown in fig. \ref{fig:betaresults}(f) and (g) respectively. It is
immediately apparent that as $N$ increases, the objective function
acquires additional local extrema. 

For the $N=1$ case, the global optimum found is $\beta_{int}=0.701632$
at $A_{0}=35.283$ and $\alpha=1.1189$, which is only marginally
short of the best known value of $\beta_{int}=0.708951$ found from
optimizing many different classes of potential \cite{Atherton2012}.
The global maximum lies at the top of the long, narrow ridge viewed
in fig. \ref{fig:betaresults}(d), a feature of the objective function
that was also seen in earlier studies of more complicated potentials
\cite{Atherton2012}. Its presence implies that $\beta_{int}$ is
much less sensitive to one of the parameters than the other, which
can be quantified by computing the hessian matrix of $\beta_{int}$
with respect to the parameters,

\[
h=\left(\begin{array}{cc}
\frac{\partial^{2}}{\partial^{2}A_{0}} & \frac{\partial^{2}}{\partial A_{0}\partial\alpha}\\
\frac{\partial^{2}}{\partial A_{0}\partial\alpha} & \frac{\partial^{2}}{\partial^{2}\alpha}
\end{array}\right),
\]
and finding the eigenvalues and eigenvectors. These quantities respectively
measure the curvature and principal directions of the objective function
around the maximum. We previously used this technique in \cite{Atherton2012}
to show that that while the best known value of $\beta_{int}=0.708951$
was obtained by optimizing a piecewise linear potential with 6 free
parameters, in fact $\beta_{int}$ was effectively only sensitive
to 2-3 parameters at the optimum. Here, the eigenvalues of the hessian
evaluated at the global maximum of $\beta_{int}$ are $-0.32277$
and $-0.00066$ and the associated eigenvectors are $(0.00965,-0.99995)$
and $(0.99995,0.00965)$. Hence, $\beta_{int}$ is very sensitive
to the value of $\alpha$, the second parameter and much less sensitive
to the value of $A_{0}$; Since the eigenvectors are nearly parallel
to basis vectors $(1,0)$ and $(0,1)$ in parameter space\textemdash as
is visible from the orientation of the ridge in fig. \ref{fig:betaresults}(d)\textemdash it
is clear that $\beta_{int}$ is sensitive to these features of the
potential specifically, and not some combination of them. In contrast,
the eigenvectors in \cite{Atherton2012} were not well aligned with
the parameter space and so it was not possible to ascribe high $\beta_{int}$
to particular features of the potential. The design advice from this
study is much clearer: to optimize $\beta_{int}$, create an asymmetric
potential well with a steep wall on one side, i.e. set $A_{0}\gg1$;
then add an attractive group in the center and tune the strength of
attraction, i.e. carefully adjust $\alpha$ as this largely determines
$\beta_{int}$. 

A similar analysis was applied to the multi-electron case. In table
\ref{tab:Beta}, we display the three extrema with largest $\beta_{int}$
for $N=8$, together with a plot of the potential and wavefunctions;
parameter values of $A_{0}$ and $\delta$ at the optimum and the
results of the eigenanalysis. Unlike the $N=1$ case, the global optimum
has a repulsive $\delta$-function; the next two solutions have attractive
$\delta$-functions. The existence of both attractive and repulsive
extrema supports the paradigm proposed by Lytel \emph{et al.} \cite{lytel2015}
in their work on optimization of quantum graphs. They suggest that
large $\beta_{int}$ is achieved by introducing a disturbance at some
point in the $\pi$-electron chain of molecule, e.g the addition of
a side group. The disturbance then induces a phase shift in the wavefunction,
producing a change in dipole moments sufficient to achieve large $\beta_{int}$.
In our work, the $\delta$-function serves to provide the disturbance;
the insight of Lytel \emph{et al.} is that it is the overall phase
shift at the disturbance that is the relevant parameter, not its detailed
nature. Hence both attractive and repulsive features can provide an
appropriate phase shift. 

Just as for the $N=1$ case above, eigenanalysis of the hessian for
the multi-electron case shows that the eigenvectors remain well aligned
with the parameter basis vectors for the multi-electron case. Surprisingly,
while $\alpha$ was found to be the most important parameter for $N=1$,
it is $A_{0}$ that appears to be most significant for the $N=8$
case. The ratio of eigenvalues is also less extreme, around $10^{-2}$
rather than $10^{-4}$ as before. These results suggest that when
applying the design approach herein proposed, i.e. an asymmetric well
with a phase shift-inducing feature, to real systems, tuning both
parameters may be important to achieve high $\beta_{int}$. 

We also performed eigenanalysis for the more general linear potential;
as a prototypical example for $N=8$ electrons the eigenvalues were
$\left(-458,-0.4,-0.1,-3\times10^{-3},1\times10^{-4},4\times10^{-6}\right)$
indicating that only one parameter is important. Interestingly, with
increasing $N$ the lowest $7$ eigenvalues remained roughly constant
while the largest eigenvalue strongly increased: For $N=2$; the principal
eigenvalue was found to be $-2.7$, for $N=4$ it was $-196$ and
for $N=6$ it was $-320$. This progression is interesting because
it suggests that for large $N$, the problem in some sense becomes
simpler as the important parameter dominates the others to an ever
increasing extent. Unfortunately, as with previous work\cite{Atherton2012},
the eigenvectors are not clearly aligned with the parameter space,
so the general linear potential provides less useful information than
the highly constrained asymmetric $\delta$-function potential.

We also display for each potential a visualization of the first few
position matrix elements $x_{nm}=\left\langle n\left|x\right|m\right\rangle $.
These are important because, as discussed more fully in the appendix
below, the hyperpolarizabilities can be expressed as a sum over states
involving $x_{nm}$ as well as the energy-level differences $E_{nm}=E_{n}-E_{m}$.
It is therefore natural to examine this matrix to determine which
transitions contribute most to the hyperpolarizabilities. The interpretation
of this matrix is, however, complicated by the fact that many combinations
of these parameters are individually irrelevant to the hyperpolarizabilities.
As is well-known, for example, the three state model\cite{kuzyk2000physical,Tripathy2004,Watkins2012}
achieves the bounds quoted in equation (\ref{eq:kuyzklimit}), and
only requires two parameters $E=\frac{E_{10}}{E_{20}}$ and $X=\frac{\left|x_{01}\right|}{\left|x_{01}^{MAX}\right|}$
with $\left|x_{01}^{MAX}\right|=\sqrt{\frac{\hbar N}{2mE_{10}}}$. 

The $x_{nm}$ matrices are dominated by the tridiagonal terms, and
the diagonal elements can be eliminated from the expressions for the
hyperpolarizabilities, e.g. by using the dipole-free sum over states
(DFSOS) formula. To aid inspection, we have therefore omitted the
diagonal terms and plotted the first off-diagonal terms, i.e. those
with $\left|n-m\right|=1$, in greyscale. The remaining terms are
plotted in a scheme where intensity corresponds to magnitude and red
or blue refers to the positive or negative sign of the term respectively.
Reflecting the potential in space and changing the signs of odd-indexed
wavefuctions changes such plots only by changing the signs, and for
$n-m\ne0$ the colors of $x_{nm}$ for $n-m$ even. To avoid confusion,
we have chosen the potential or its mirror image in such a way that
$x_{nm}$ is positive for $n$ the HOMO and $m$ the LUMO+1. The off-tridiagonal
terms are important to $\beta_{int}$, because a tridiagonal matrix
would yield $\beta_{int}=0$ as for the Harmonic Oscillator. Clearly,
however, for all the local optima displayed in table \ref{tab:Beta},
the $\left|n-m\right|=2$ terms are much larger than the remaining
$\left|n-m\right|>2$ terms. 

For each local maximum in table \ref{tab:Beta}, we display calculated
values of the $E$ and $X$ parameters. For the three state model,
these parameters yield optimal $\beta_{int}=1$ for values of $E=0$
and $X=3^{-1/4}=0.760$. However, past studies\cite{Zhou2007} of
optimized potential functions for $N=1$ electrons find values of
$E\approx1/2$ and $X\mbox{\ensuremath{\approx}}0.789$ regardless
of the starting potential. Optimization of Quantum graphs\cite{lytel2013dressed}
produces mildly different values of $E\approx0.4$ and $X\approx0.79$.
For the optima presented here, we also find $E=1/2$, but the results
seem to favor a value of $X=0.57$. This result is consistent with
the results of eigenanalysis of the hessian, which suggests only one
of the parameters $E$ and $X$ can be important. 

The dipole matrix plots, together with the value of $X$ or $X'=x_{HOMO,LUMO+1}/x_{HOMO,LUMO}$
allows us to appreciate the compromises made in this optimization.
All contributions to $\beta$ in the dipole-free SOS formula involve
three states, the product of the transition moments between them and
a function of the energies. From the x-matrix plots it is clear that
matrix elements become smaller very quickly moving away from the diagonal.
Thus the SOS for $\beta$ is expected to be dominated by contributions
from terms that involve only one off-tridiagonal element. There are
exactly two sets of three states involving only the one off-tridiagonal
matrix element: the ground state, the excited state in which one electron
has been excited from the HOMO to the LUMO and one of two doubly excited
states: the state in which one electron has been excited from the
HOMO to the LUMO+1 or that in which one electron has been excited
from the HOMO-1 to the LUMO. As the HOMO to LUMO+2 and HOMO-1 to LUMO
transition matrix elements have opposite signs, these contributions
have opposite signs. Generically the transition moments are larger
between higher energy states to the HOMO to LUMO+1 and so are expected
not to dominate. 

The smaller result for many electrons seems, at least in part, to
be explained by the negative contribution of the HOMO-1 to LUMO contribution.
This contribution is lacking for 1 or two electrons when there is
no HOMO-1. Other than this, the pattern of matrix elements looks substantially
similar for all the maxima\textemdash including ones we have not displayed.
Thus it seems that the smaller results for $\beta_{int}$ for more
than two electrons may be explained by the need to include four states
rather than two, given the near degeneracy of the two doubly excited
states.

\subsection{Second hyperpolarizability}

\begin{figure}
\begin{centering}
\includegraphics[width=3in]{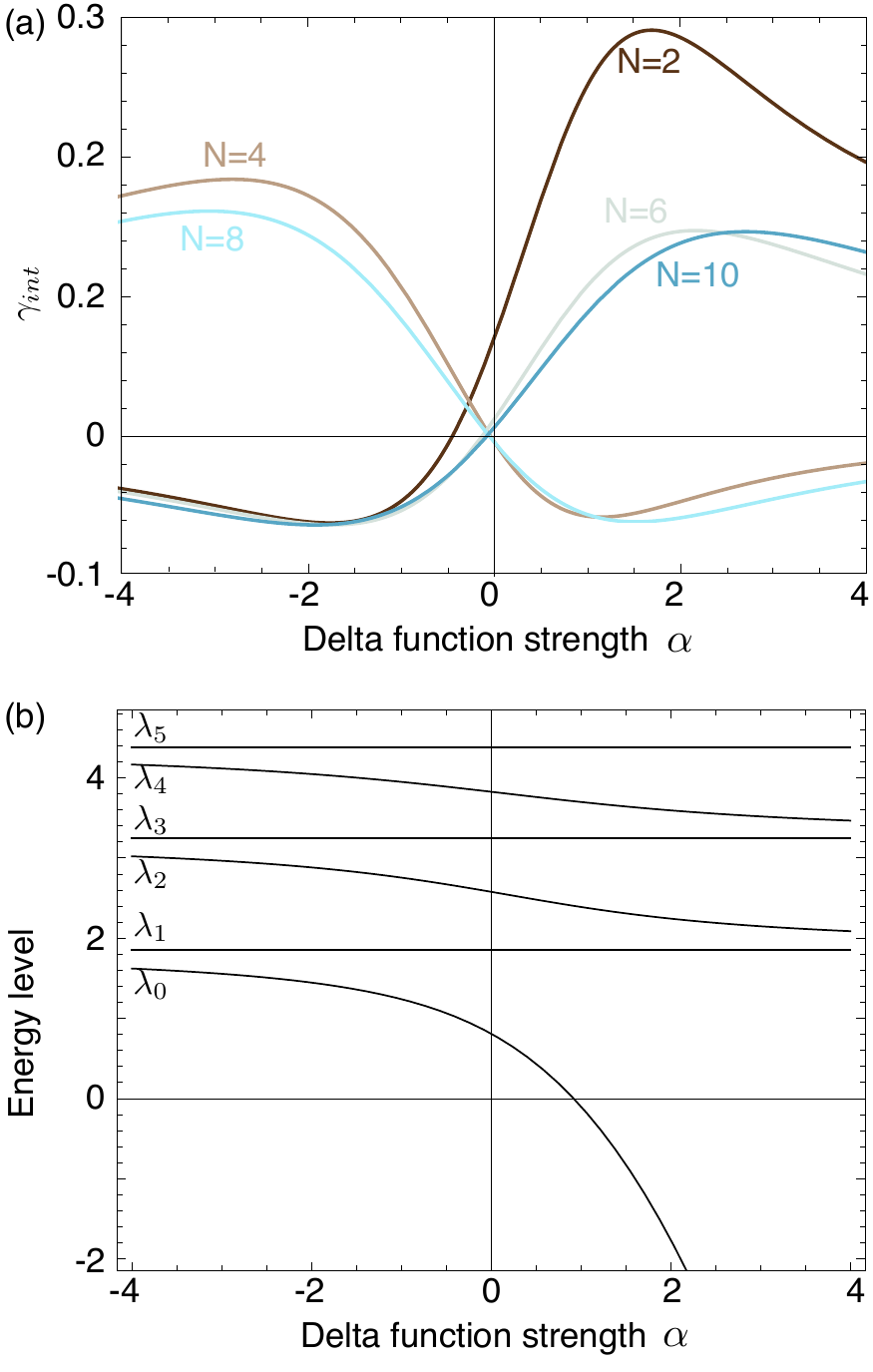}
\par\end{centering}

\caption{\label{fig:gammadelta}(a) $\gamma_{\mathrm{int}}$ versus $\delta$
function strength $\alpha$ for varying electron number $N$ in the
symmetric $\delta$ function potential. Each time $N$ is increased
by 2, the sign of $\alpha$ for which the maximum or minimum occurs
changes sign. (b) The energy level structure of the symmetric $\delta$
function potential as a function of $\delta$ function strength $\alpha$. }
\end{figure}

The symmetric triangular well with a $\delta$-function, i.e. eq.
(\ref{eq:potentialdelta}) with $A_{0}=1$, is known to achieve near-optimal
results for $\gamma_{int}$ for $N=1$ despite only containing one
free parameter. Due to the simplicity of this potential, it is possible
to directly visualize $\gamma_{int}$ as a function of this parameter,
the strength of the delta function $\alpha$, for different values
of $N$. The results are displayed in fig. \ref{fig:gammadelta}(a).
Immediately apparent is an alternation of the sign of the $\gamma_{int}$
curves as $N$ is increased by 2: for $N=2,6,10...$ the maximum value
of $\gamma_{int}$ occurs for positive $\alpha$\textemdash corresponding
to a $\delta$-function wall in the middle of the well, while for
$N=4,8...$ the maximum value of $\gamma_{int}$ occurs for negative
$\alpha$. The magnitude of $\alpha$ for which the maxima occur increases
only slightly with $N$, however. The behavior of the minimum value
of $\gamma_{int}$ is similar, but shifts even less. 

Some insight into these results is obtained by examining the effect
of the $\delta$-function on the single electron energy level spectrum
for the potential, shown in fig. \ref{fig:gammadelta}(b). Clearly,
only the even wavefunctions are affected by the $\delta$-function
due to the $\mathcal{P}$-symmetry of the potential. While the placement
of $\lambda_{0}$ can be freely adjusted by changing $\alpha$, the
higher even energy levels are bounded by their invariant odd neighbors,
e.g. $\lambda_{1}<\lambda_{2}<\lambda_{3}$. The freedom of selecting
$\lambda_{0}$ relative to $\lambda_{1}$ allows the $N=2$ case to
achieve $\gamma$ a high fraction of the Kuzyk maximum, while the
restricted higher levels only permit a lower value of $\gamma_{int}$
to be achieved. The reason for the alternation is also apparent. For
$N=2$, the Highest Occupied Molecular Orbital (HOMO) in the ground
state is $\lambda_{0}$ while the Lowest Unoccupied Molecular Orbital
(LUMO) is $\lambda_{1}$; increasing $\alpha$ serves to \emph{widen}
the HOMO-LUMO gap. On the other hand, for $N=4$ the HOMO is $\lambda_{1}$
and the LUMO is $\lambda_{2}$; increasing $\alpha$ for this case
serves to narrow the HOMO-LUMO gap. This trend continues with $\alpha$
widening the HOMO-LUMO gap for $N=2,6,10...$ and narrowing the HOMO-LUMO
gap for $N=4,8$; the alternating effect of $\alpha$ explains the
different signs of $\gamma_{int}$ with $N$.

\noindent 
\begin{figure}
\begin{centering}
\includegraphics[width=1\columnwidth]{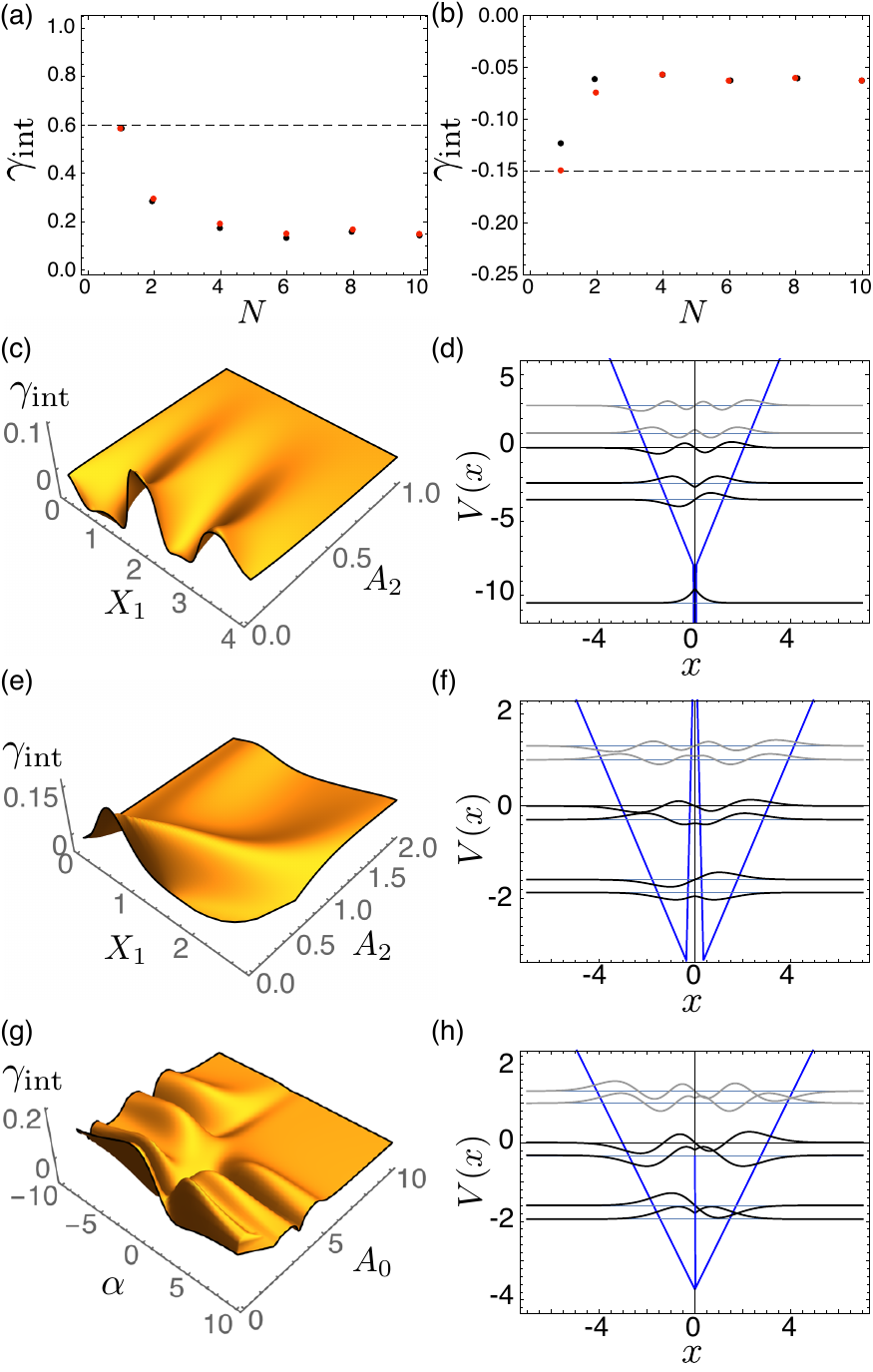}
\par\end{centering}

\caption{\label{fig:gammaall}Maximum (a) and minimum (b) $\gamma_{\mathrm{int}}$
achieved for varying increasing numbers of electrons; results for
the symmetric linear piecewise potential are shown in red and for
the symmetric $\delta$ function potential in black. (c) For $N=8$,
the objective function $\gamma_{\mathrm{int}}$ plotted versus the
linear piecewise potential shape parameters $A_{2}$ and $X_{1}$
for positive $A_{1}$ and (d) the best potential and wavefunctions
obtained. (e) Results for negative $A_{1}$ and (f) the corresponding
potential and wavefunctions. (g) Results for the $\delta$ function
potential shape parameters $\alpha$ and $A_{2}$ and (h) optimized
wavefunctions and potential. }
\end{figure}

To determine whether these results are universal, we also optimized
$\gamma_{int}$ for the asymmetric $\delta$-function potential eq.
(\ref{eq:potentialdelta}) as well as the linear piecewise potential
given by eq. (\ref{eq:potentiallinear}). Shown in fig. \ref{fig:gammaall}(a)
and (b) are the best maximum and minimum $\gamma_{int}$ obtained
as a function of $N$. Although there are small differences between
results obtained with different potentials, the same trend is clear:
that the best $\gamma_{int}$ falls off with increasing $N$ but rapidly
reaches a constant value, yielding an apparent feasible range of $-0.05<\gamma_{int}^{max}<0.2$.
These apparent bounds are shared by all three potentials.

We plot the objective function results for $N=8$ for several different
scenarios: For the linear piecewise potential, the sign of $A_{1}$
must be chosen to be positive or negative prior to optimization. The
objective function is shown for $A_{1}=+1$ in fig. \ref{fig:gammaall}(c)
revealing several local maxima; the corresponding optimized potential,
which maximizes $\gamma_{int}$ , and wavefunctions are shown in fig.
\ref{fig:gammaall}(d). If alternatively, $A_{1}=-1$, a minimum of
$\gamma_{int}$ is obtained; the objective function is rather simpler
as apparent in fig. \ref{fig:gammaall}(e) and the corresponding potential
and wavefunctions are shown in fig. \ref{fig:gammaall}(f). Note that
despite the arbitrariness of the linear potential, the optimized potentials
strongly resemble the symmetric $\delta$-function potential, validating
its use as \emph{ansatz} earlier. 

\begin{table*}
\begin{centering}
\begin{tabular}{>{\centering}m{0.5in}>{\centering}m{1in}>{\centering}m{0.5in}>{\centering}m{0.5in}>{\centering}m{0.75in}>{\centering}m{0.75in}>{\centering}m{1in}>{\centering}m{0.5in}>{\centering}m{0.75in}}
\hline 
Value of $\gamma_{int}$ & Potential and Wavefunctions & $A_{0}$ & $\alpha$ & Hessian eigenvalues & Hessian eigenvectors & $x_{nm}$ & $X$ & $E$\tabularnewline
\hline 
\hline 
0.161 (100\%) & \includegraphics[width=1in]{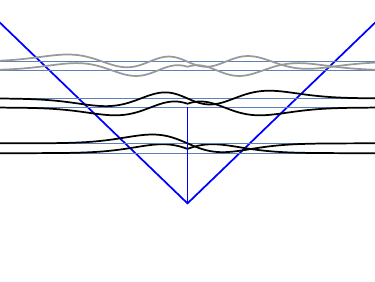} & 1.000 & -2.963 & -0.597, -0.048 & $\left(\begin{array}{c}
-0.999\\
0.053
\end{array}\right)$ & \includegraphics[width=1in]{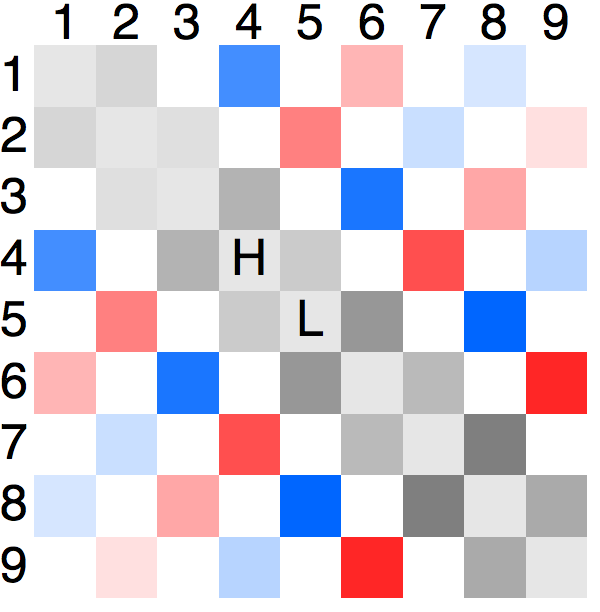} & 0.615 & 0.765\tabularnewline
0.134 (83\%) & \includegraphics[width=1in]{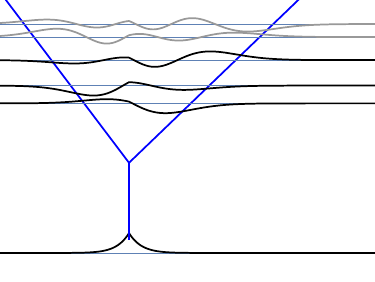} & 1.375 & 2.534 & -1.926, -0.039 & $\left(\begin{array}{c}
-0.999\\
-0.004
\end{array}\right)$ & \includegraphics[width=1in]{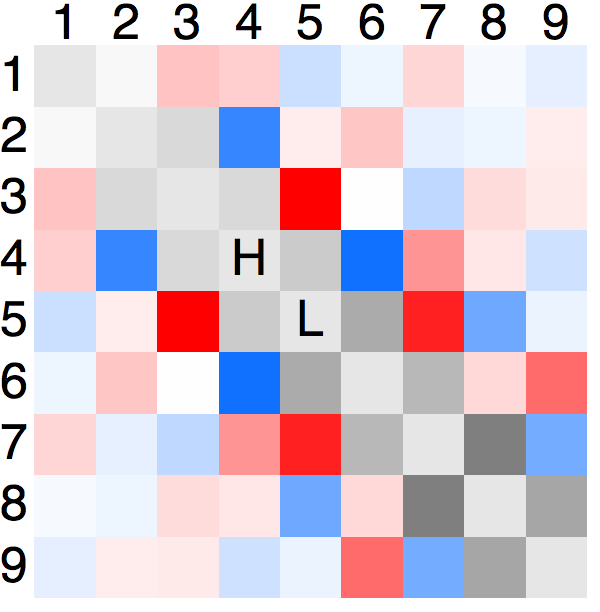} & 0.516 & 0.638\tabularnewline
0.127 (79\%) & \includegraphics[width=1in]{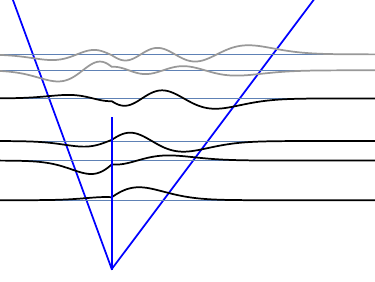} & 2.043 & -3.765 & -1.597, -0.039 & $\left(\begin{array}{c}
-0.999\\
0.024
\end{array}\right)$ & \includegraphics[width=1in]{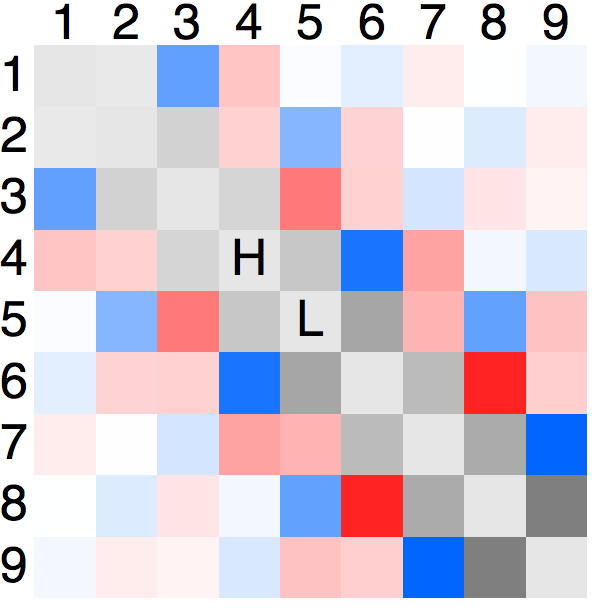} & 0.511 & 0.638\tabularnewline
\hline 
-0.0613 (100\%) & \includegraphics[width=1in]{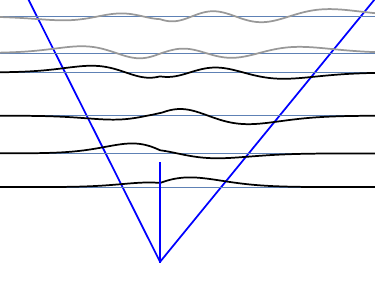} & 1.630 & -2.107 & 6.970, 0.0187  & $\left(\begin{array}{c}
0.999\\
-0.016
\end{array}\right)$ & \includegraphics[width=1in]{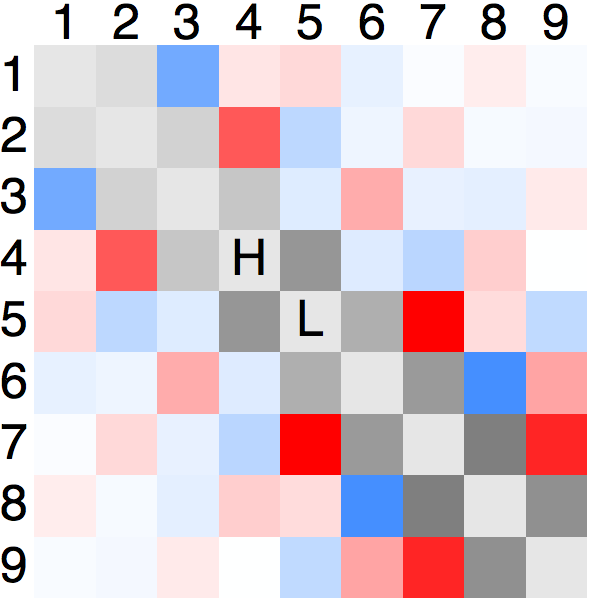} & 0.655 & 0.500\tabularnewline
-0.0609 (99\%) & \includegraphics[width=1in]{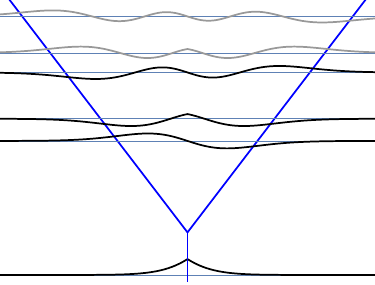} & 1.000 & 1.524 & 13.78,

0.0311 & $\left(\begin{array}{c}
0.999\\
-0.0006
\end{array}\right)$ & \includegraphics[width=1in]{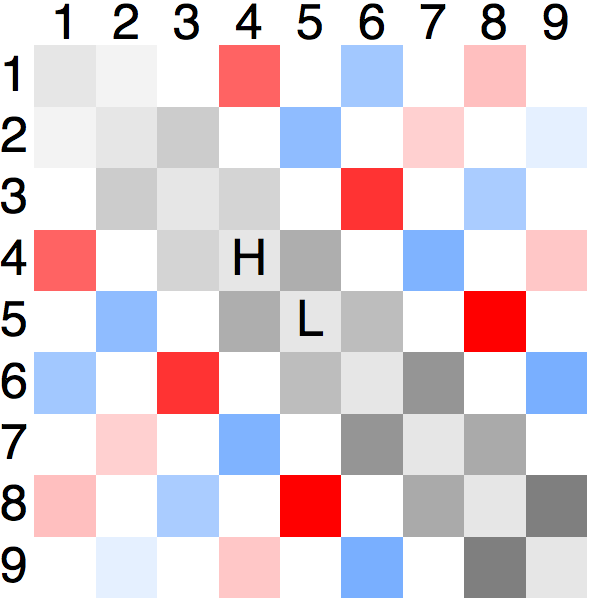} & 0.654 & 0.500\tabularnewline
-0.0502 (82\%) & \includegraphics[width=1in]{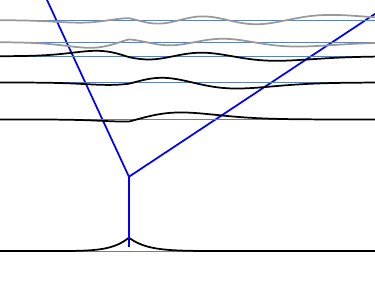} & 3.260 & 2.467 & 1.162, 0.0123,  & $\left(\begin{array}{c}
0.997\\
-0.069
\end{array}\right)$ & \includegraphics[width=1in]{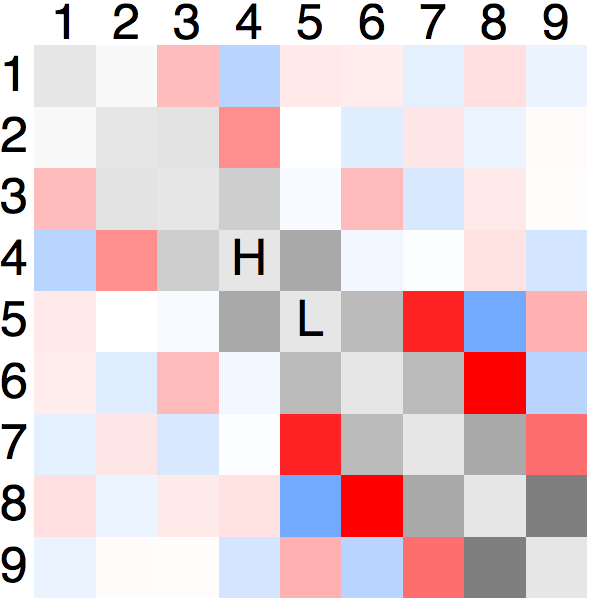} & 0.639 & 0.500\tabularnewline
\hline 
\end{tabular}
\par\end{centering}

\caption{\label{tab:AsymmetricGamma}Asymmetric $\delta$-function potentials
with locally optimal $\gamma_{int}$. Results for each potential are
shown as follows, columnwise: the value of $\gamma_{int}$ , and the
fraction of the globally optimum value; the potential and wavefunctions;
optimized values of the parameters $A_{0}$ and $\delta$; the eigenvalues
and eigenvectors of the Hessian matrix of $\gamma_{int}$ with respect
to the parameters; a plot of the dipole transition matrix $x_{nm}=\left\langle n\left|x\right|m\right\rangle $
with the HOMO and LUMO indicated by H and L respectively and coloring
as described in the text; the energy ratio $E$ and dipole transition
moment $X$ for the Kuzyk three-state model. }
\end{table*}

The asymmetric $\delta$-function potential has a much more complicated
landscape for $\gamma_{int}$ as is evident in fig. \ref{fig:gammaall}(g).
Many local minima and maxima exist. The best three for both positive
and negative $\gamma$ are displayed in table \ref{tab:AsymmetricGamma},
together with the parameter values and results of the hessian eigenanalysis.
Many of the results above for $\beta$ are also seen for $\gamma$:
both attractive and repulsive $\delta$-functions lead to large $\gamma$,
in agreement with the phase interruption paradigm. Eigenvalues of
the hessian differ by a similar ratio of $\sim10^{-2}$ and again
the eigenvectors are well aligned with the basis vectors of the parameter
space. Just as for the first hyperpolarizability with $N=8$ electrons,
it appears that $A_{0}$ is the most important parameter. 

Intriguingly, the best negative $\gamma_{int}$ obtained does not
possess $\mathcal{P}$ symmetry in sharp contrast to the $N=1$ case,
although a $\mathcal{P}$-symmetric solution within 1\% of this value
also exists. It is possible this is due to the freedom of choosing
the relative position of the lowest energy level $\lambda_{0}$ relative
to $\lambda_{1}$; less freedom exists for the higher single particle
energy levels.

Position matrix elements $x_{nm}$ are shown in \ref{tab:AsymmetricGamma}
for each potential. Note that these matrices for $\mathcal{P}$ symmetric
potentials have a characteristic checkerboard structure. Elements
immediately off the diagonal $\left|n-m\right|=1$ are plotted on
a greyscale while non-tridiagonal elements are plotted in red and
blue and it is evident, as for $\beta_{int}$, that the matrices are
diagonally dominant. For positive $\gamma_{int}$ , the $\mathcal{P}$-symmetric
global optimum has large terms on the $|n-m|=3$ off diagonal, which
is consistent with a few-state hypothesis; here the HOMO$\to$LUMO+2
and HOMO-1$\to$LUMO+1 terms make the strongest contribution. The
transition matrices for the non $\mathcal{P}$-symmetric solutions,
however, seem to have little in common with the global optimum, which
is perhaps to be expected since the values of $\gamma_{int}$ for
these secondary maxima fall some $20\%$ below it. 

For negative $\gamma_{int}$, where the two best solutions have nearly
identical values, the transition matrix for the $\mathcal{P}$-symmetric
optimum resembles the solution for positive $\gamma_{int}$, i.e.
HOMO$\to$LUMO+2 and HOMO-1$\to$LUMO+1 are the dominant off-tridiagonal
terms. The non $\mathcal{P}$-symmetric solution has an interesting
structure: the HOMO$\to$LUMO element is large, while the HOMO$\to$LUMO+1
and HOMO-1$\to$LUMO elements are significantly weaker than the other
$\left|n-m\right|=2$ elements. A similar pattern, where off-tridiagonal
elements are suppressed around the Fermi surface, is also seen in
the third highest optimum that is also non $\mathcal{P}$-symmetric.
It is for negative $\gamma_{int}$, therefore, that we find the clearest
correspondence with the transition matrix elements of the three-state
ansatz. 

We now connect these results to the Sum-Over States Formula for $\gamma$,
which includes two types of terms: There are both terms involving
the ground state and three other states, $m$, $n$, and $l$, energy
factors and the matrix element product $x_{0l}x_{0n}x_{lm}x_{nm}$
or $x_{0m}x_{ml}x_{ln}$ which we shall hereafter call ``four state''
terms. There are also terms involving the ground state and two excited
states, $n$, $m$, referred to hence as ``three state'' terms.
A three state term proportional to $(x_{0n}x_{nm})^{2}$ makes a positive
contribution to $\gamma$, one proportional to $(x_{0n}x_{0m})^{2}$
will make a negative contribution. This is more easily seen from eq.
(73) in \cite{kuzykreview}, if $m$ is considered the unconstrained
summation variable, $n$ and $l$ are (as before) both considered
to be different from $m$, and the energy factors are made symmetric
in $n$ and $l$. Moreover, only three state terms involving at least
one off-tri-diagonal matrix element can contribute. If there are no
off-tri-diagonal matrix elements, the matrix elements and energies
are constrained to be those for the harmonic oscillator, for which
$\gamma=0$. The largest terms in $\gamma$ from four-state terms
involves three excited states, three-tridiagonal matrix elements and
one matrix element for which $|n-m|=3$. The largest contributions
to $\gamma$ from three state terms involves the square of a tri-diagonal
matrix element and the square of an $|n-m|=2$ matrix element. These
last matrix elements, and so also these terms, are forbidden by $\mathcal{P}$
symmetry.

Now looking at the $x$ matrices, we see that the the $\mathcal{P}$-symmetric
potentials have three large contributions, with the signs of these
three contributions oscillating with period 2 in the empty state in
the highest or lowest singly occupied state. The resultant cancellations
may again partly explain the smaller $\gamma_{int}$ achieved for
these potentials. The difference between positive and negative $\gamma$
is not the kinds of terms involved (as suggested in \cite{Kuzyk00})
but rather in the signs of the off-tridiagonal matrix elements. The
three-state terms that are allowed for non $\mathcal{P}$-symmetric
potentials are more similar to those considered in \cite{Kuzyk00},
except that these can have either sign.

To facilitate comparisons with other work, we again computed values
of the Kuzyk three-state model parameters $E$ and $X$. For the lower
bound, all three extrema favor $E=0.5$ and $X\approx0.65$. An earlier
study by Watkins \emph{et al.}\cite{Watkins2012}, who optimized $\gamma_{int}$
for different potentials and $N=1$, found for negative $\gamma_{int}$
$X\approx0.66$ and two clusters of solutions with $E\approx0.15$
or $E\approx0.53$. Our results are therefore in excellent agreement.
For the upper bound, our results display more variation with $E=0.765$
and $X=0.615$ for the best potential found. Watkins \emph{et al.}
similarly found more variation in positive $\gamma_{int}$, obtaining
results around $E\approx0.57$ with $X\approx0.5$. As for $\beta$,
the discrepancy is explicable because eigenanalysis reveals that effectively
only one parameter is truly important, seemingly the slope of the
left hand boundary.

\subsection{Discussion\label{sec:Discussion}}

\begin{figure}
\includegraphics[width=1\columnwidth]{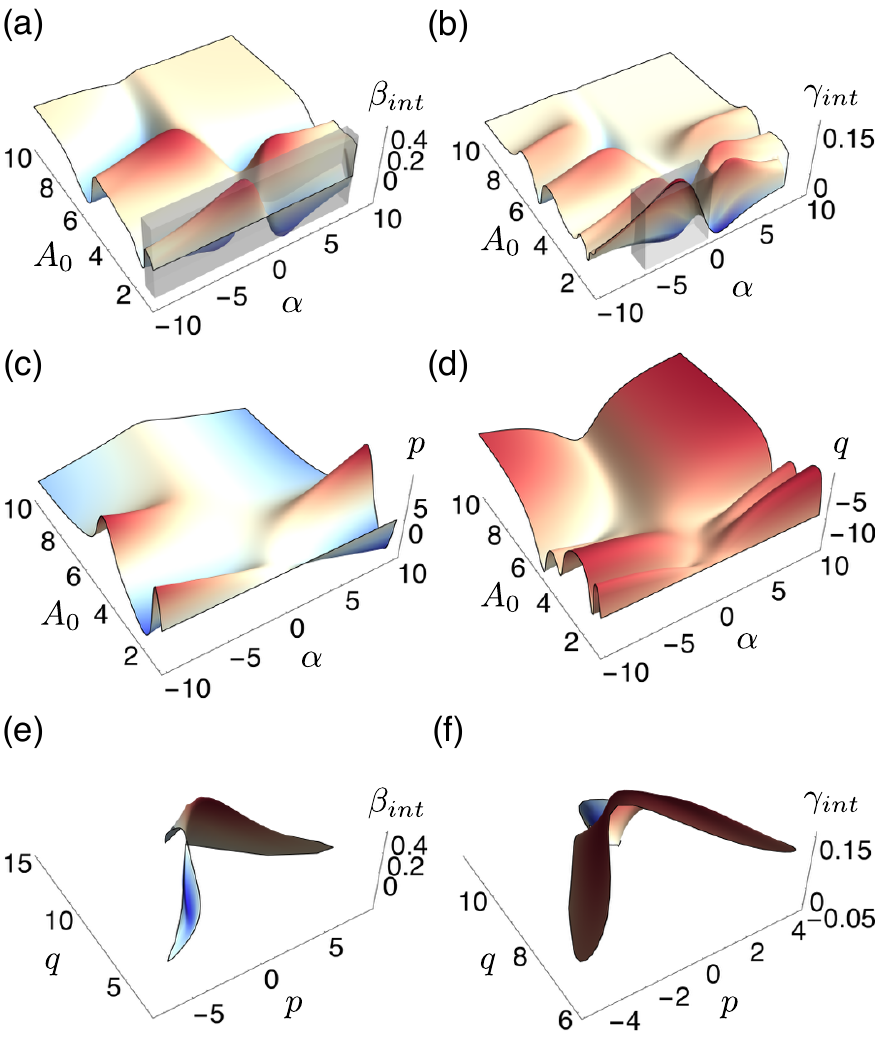}

\caption{\label{fig:Reparam}Objective functions (a) $\beta_{int}$ and (b)
$\gamma_{int}$ for the asymmetric $\delta$-function potential shown
in the natural parameter space $A_{0}$ and $\alpha$; wavefunction
parameters at the classical turning points (c) $p=\psi_{l}'/\psi_{r}'-\psi_{r}'/\psi_{l}'$
and (d) $q=-\Delta x^{2}\left|\psi'_{l}\psi'_{r}\right|^{2/3}$ plotted
with respect to $A_{0}$ and $\alpha$. A selected region of the objective
function, enclosed in the grey volume in (a) and (b), is show replotted
with respect to $p$ and $q$ in (e) and (f) respectively. }
\end{figure}

The above subsections have presented optimized potentials for $\beta_{int}$
and $\gamma_{int}$ with increasing $N$, expressed in a parameter
space that \emph{a posteriori} is found to coincide with what is important
to the hyperpolarizabilities. As discussed, this is a significant
development over previous work, where the parameter space was larger
the optimized potentials gave a less clear sense of which features
are important. Nonetheless, the parameters $A_{0}$ and $\alpha$
are an artifact of the parameterization chosen, and it is therefore
desirable to identify quantities that are invariant under reparametrization
and rescaling of the form in eq. (\ref{eq:rescaling}). This is particularly
important because, if we generalize to the space of arbitrary potentials,
there are many other solutions with identical values of the hyperpolarizability.
For the $N=1$ case, we know that the hyperpolarizabilities can be
completely reconstructed from the ground state wavefunction\cite{Wiggers07}.
Extending this to the $N>1$ case, it is easy to see that the hyperpolarizablities
can be constructed for this problem as appropriate integrals of the
occupied single particle wavefunctions alone. As the largest term
is expected to the HOMO, the important parameters ought to be expressible
in terms of parameters crucial to the HOMO.

Moreover, there is a need to connect the dipole-free sum-over-states
view, which prescribes certain values of the dipole transition matrix
elements, and the potential view that we pursued above. We therefore
performed an approximate analysis of the problem, inserting WKB ansatz
wavefunctions into expressions for $E_{nm}$ and $x_{nm}$ to determine
what details of these wavefunctions most significantly affect these
parameters. The full calculations presented in the Appendix below
support this argument and suggest that the three parameters introduced
in the next paragraph are the crucial ones.. 

On the basis of approximate analytic arguments made in the appendix,
and a numerical experiment that follows, we believe that there are
three length scales that are important to the hyperpolarizabilities
around the maximum. One is $\Delta x$, the separation between the
turning points of the HOMO. The other two are length scales $L_{k}$
for $k$ and an element of $\{r,l\}$ that characterizes the rate
at which the wavefunction varies near the right and left turning points
of the HOMO. A variety of nearly equivalent forms can be given for
these lengths and dimensionless ratios containing them, including
$L=(V')^{-1/3}$, and more global formulae related the derivatives
of the wavefunction at the turning points or to the normalization
of the HOMO. From these three lengths, it is possible to construct
two dimensionless parameters that are arguably important to the hyperpolarizability,
for example, 
\begin{eqnarray}
p & = & \psi_{l}'/\psi_{r}'-\psi_{r}'/\psi_{l}'\nonumber \\
q & = & -\Delta x^{2}\left|\psi'_{l}\psi'_{r}\right|^{2/3}\label{eq:newparams}
\end{eqnarray}
are invariant under reparameterization and rescaling. Here $\Delta x$
is the distance between the turning points, $\psi_{l}'$ and $\psi_{r}'$
are the slopes of the wavefunctions where the subscripts $l$ and
$r$ refer to the left and right turning point respectively. As some
of the definitions of $L$ are zero when $V'$ is infinite at the
turning points, and it is still possible for the hyperpolarizabilities
to approach the maximum in this case, appropriate dimensionless combinations
should not be infinite in this limit. As $\beta_{int}$ requires asymmetry
in the potential, it is expected that a parameter odd under $\mathcal{P}$
symmetry, such as $p$, mostly controls $\beta$ while one even under
$\mathcal{P}$ symmetry, such as $q$, mostly controls $\gamma_{int}$.

In fig. \ref{fig:Reparam} we show the objective functions (a) $\beta_{int}$
and (b) $\gamma_{int}$ plotted in the parameter space of the asymmetric
$\delta$-function potential, i.e. $A_{0}$ and $\alpha$. In the
same parameter space, we show in \ref{fig:Reparam}(c) and (d) the
values of the wavefunction parameters $p$ and $q$ from eq. (\ref{eq:newparams}).
From these plots, we see that much of the structure in the objective
functions $\beta_{int}$ and $\gamma_{int}$ is attributable to these
new parameters: notice that the position of the ridges in \ref{fig:Reparam}(c)
corresponds to the ridges in \ref{fig:Reparam}(a), while those in
\ref{fig:Reparam}(d) corresponds, more roughly, those in \ref{fig:Reparam}(b).
In figs. \ref{fig:Reparam}(e) and (f), the portion of the objective
function in figs. \ref{fig:Reparam}(a) and (b) that is enclosed within
the grey cuboid is reprojected into the new ($p$, $q$) parameter
space. Plots of the objective functions\textemdash not shown here\textemdash in
$(p,q)$ space for the full range of $(A_{0},\alpha)$ from fig. \ref{fig:Reparam}(a)
and (b) closely resemble the structure observed in this reduced reduced
region.

From these results, we conclude that $\beta_{int}$ is largely determined
by $p$ while $\gamma_{int}$ is largely determined by $q$, and the
remaining parameter in each case must be tuned less precisely to achieve
the optimum. It is also now clear that the asymmetric $\delta$-function
performs so well due to a fortuitous correspondence: the $A_{0}$
parameter directly controls the ratio of the slope of the potentials
at the turning points, which is readily related to the slope of the
wavefunctions at the classical turning points, i.e $A_{0}$ and $p$
are simply related. 

\begin{table}
\begin{centering}
\begin{tabular}{|c|c|c|c|}
\hline 
 & Optimum value & $p$ & $q$\tabularnewline
\hline 
\multirow{3}{*}{$\beta_{int}$} & $0.403$ & $1.69$ & $-8.99$\tabularnewline
 & $-0.372$ & $-1.28$ & $-7.66$\tabularnewline
 & $0.370$ & $2.11$ & $-9.26$\tabularnewline
\hline 
\multirow{3}{*}{$\gamma_{int}$$>0$} & $0.161$ & $0$ & $-9.30$\tabularnewline
 & $0.134$ & $-1.85$ & $-6.40$\tabularnewline
 & $0.126$ & $-1.88$ & $-6.95$\tabularnewline
\hline 
\multirow{3}{*}{$\gamma_{int}<0$} & $-0.0613$ & $0.456$ & $-10.54$\tabularnewline
 & $-0.0610$ & $-1.02\times10^{-5}$ & $-9.30$\tabularnewline
 & $-0.0503$ & $1.214$ & $-10.54$\tabularnewline
\hline 
\end{tabular}
\par\end{centering}

\caption{\label{tab:Dimensionlessparams}Values of dimensionless parameters
$p$ and $q$ constructed from the HOMO wavefunction for each of the
optimized $N=8$ solutions displayed in tables \ref{tab:Beta} and
\ref{tab:AsymmetricGamma}}
\end{table}

\begin{table}
\begin{centering}
\begin{tabular}{|c|c|c|c|c|c|c|}
\cline{3-7} 
\multicolumn{1}{c}{} &  & \multicolumn{5}{c|}{$N$}\tabularnewline
\cline{3-7} 
\multicolumn{1}{c}{} &  & $2$ & $4$ & $6$ & $8$ & $10$\tabularnewline
\hline 
\multirow{2}{*}{$\beta_{int}$} & $p$ & $-4.17$ & $1.94$ & $-2.52$ & $1.69$ & $-1.96$\tabularnewline
 & $q$ & $-1.37$ & $-4.00$ & $-7.11$ & $-8.99$ & $-12.6$\tabularnewline
\hline 
\multirow{2}{*}{$\gamma_{int}$$>0$} & $p$ &  & $5.15\times10^{-3}$ & $2.08$ & $0$ & $1.92$\tabularnewline
 & $q$ &  & $-3.64$ & $-4.97$ & $-9.30$ & $-9.62$\tabularnewline
\hline 
\multirow{2}{*}{$\gamma_{int}<0$} & $p$ &  & $0$ & $0$ & $0.456$ & $0$\tabularnewline
 & $q$ &  & $-3.65$ & $-7.35$ & $-10.5$ & $-13.2$\tabularnewline
\hline 
\end{tabular}
\par\end{centering}

\caption{\label{tab:DimensionlessFunctionN}Dimensionless parameters $p$ and
$q$ calculated from the best solution found as a function of $N$. }
\end{table}

We display in table \ref{tab:Dimensionlessparams} the values of $p$
and $q$ calculated for each of the optima of $\beta_{int}$ and $\gamma_{int}$
shown in table \ref{tab:Beta} and \ref{tab:AsymmetricGamma}. For
each of the optimization problems, i.e. $\left|\beta_{int}\right|$,
$\gamma_{int}>0$ and $\gamma_{int}<0$ the values of at least one
of these parameters are internally quite consistent with each other.
This is particularly so for the negative $\gamma_{int}$ results where
the secondary optima are close to the global optimum. Values of $p$
and $q$ for the global optimum as a function of $N$ are shown in
table \ref{tab:DimensionlessFunctionN}. These reveal several trends:
First $\text{sgn}(\beta_{int})=\text{sgn}(p)(-1)^{n_{nh}}$ where
$n_{nh}$ is the number of nodes in the HOMO. Otherwise, it is expected
that $\mathcal{P}$ symmetry just changes the sign of $\beta_{int}$
and $p$. Also, a sign alternation occurs in $p$ for $\beta_{int}$;
$q$ seems to increase with $N$ for $\gamma_{int}$, while $p$ is
found alternately $0$ or some small value. The \emph{a posteriori}
consistency of these parameters supports the argument above that these
are the ``real'' parameters of the optimization problem.

\section{Conclusion\label{sec:Conclusion}}

We have optimized the intrinsic hyperpolarizabilities $\beta_{int}$
and $\gamma_{int}$ for non-interacting multi-electron systems with
respect to the shape of several classes of potential: a piecewise
linear potential, and an asymmetric triangular well with a $\delta$-function.
The best values obtained for $\beta_{int}$ and $\gamma_{int}$ drop
from the $N=1$ case and approach an apparent feasible range of $\left|\beta_{int}\right|<0.4$
and $-0.05<\gamma_{int}^{max}<0.2$ for $N$ larger than around eight
electrons. The asymmetric $\delta$-function potential achieves these
bounds and, due to the small number of parameters, and \emph{a posteriori}
verification that the parameters are indeed relevant, provides a design
prototype for synthesis of new chromophores. For $\beta_{int}$, a
molecule should have asymmetric walls and possess an attractive or
repulsive feature in the middle\textemdash a main chain functional
group or side-chain\textemdash that promotes a rapid change in the
phase of the wavefunction. The asymmetry of the boundary and the strength
of the attraction or repulsion should then be tuned to achieve high
$\beta_{int}$. For $\gamma_{int}$, the molecule should be essentially
$\mathcal{P}$-symmetric with a central attractive or repulsive feature
that should similarly be tuned. 

By approximate analysis, we also determined that the \emph{ad hoc}
parameters of our potential can be re-expressed in terms of the dimensionless,
scale invariant parameters derived from the shape of the HOMO wavefunction
at the classical turning points. These new parameters are important
both because they explain the success of our original parametrization
and because they provide a new wavefunction-centered approach to screen
potential chromophores for large hyperpolarizabilities. 

The results also provide important information on how well the many
theoretical studies of single electron systems might describe real
molecules that posses multiple electrons. While the apparent bounds
quoted above are more restrictive than the $N=1$ case, it seems,
encouragingly, that the overall design paradigms described above apply
equally well to both cases. Information from other studies on target
values of the $E$ and $X$ parameters of the three state model for
$N=1$ also seems to remain valid with increasing $N$. The insights
of \cite{Atherton2012,burke2013} that only a very small number of
parameters are required remain valid, and happily, this appears to
be increasingly so with large $N$. Because of this and because the
objective function for multiple electrons acquires many more local
extrema, it should in principle be \emph{easier} to tune multi-electron
systems. 

\appendix

\section{Approximate Analysis}

In this appendix we use approximate techniques to argue that the dimensionless
parameters identified in the main text constructed from the separation
between the turning points of the HOMO and the slopes of the potential
at these turning points ought to mostly explain the hyperpolarizability.
The simple, zeroth order argument is given in the next two paragraphs.
A more detailed but still very approximate calculation follows.

For all potentials with large hyperpolarizabilities, the transition
matrix elements between the frontier wavefunctions\textemdash those
that are close to the Fermi level\textemdash must be large. When these
matrix elements are large, there are very strong constraints on the
energy differences between the various states, and hence we can regard
these parameters are fixed by the matrix elements and we need only
calculate the transition matrix elements to predict the hyperpolarizabilities. 

In order to calculate these transition matrix elements approximately,
we note that wavefunctions fall rapidly in the classically forbidden
region and oscillate in the classically allowed region. Thus most
of the transition matrix integral between two states necessarily comes
from the region between or very close to the turning point of the
lower energy state. Moreover, far from the turning points, each wavefunction
becomes relatively small, in effect because the particles are (classically)
moving relatively quickly. Thus most of the contribution to these
integrals comes from the region close to the turning point of the
lower energy wavefunction. From this it is clear that the distance
between the turning points of the HOMO is an important parameter.

The wavefunction near a turning point is constrained by the slope
of the potential at the turning point. Moreover, the ratio of the
amplitudes of the wavefunction at the two turning points is limited
by general principles, described below. While the wavefunction of
the higher energy state near the turning points of the lower energy
states is more free to vary, provided the potential is not too far
from linear, they are still largely constrained by the slope of the
potential. Thus the slopes of the potential at the turning points
of the HOMO also seems to be a very important parameter. From these
parameters, it is possible to construct two dimensionless parameters,
and it is known from numerical experiments that only two parameters
seem to be important to maximizing hyperpolarizabilities for model
potentials.

We now provide a more detailed, but nonetheless approximate, analysis.
We begin with the sum-over-states formulae for the hyperpolarizabilities.
For instance, the off-resonant expression for the second hyperpolarizability
is,\begin{widetext}
\begin{equation}
\gamma_{xxxx}=2e^{4}\left(2\sum_{lmn}^{\infty}{}^{'}\frac{x_{0l}\bar{x}_{lm}\bar{x}_{mn}x_{n0}}{E_{l0}E_{m0}E_{n0}}-\sum_{mn}^{\infty}{}^{'}x_{0m}^{2}x_{0n}^{2}\left\{ \frac{1}{E_{m0}^{2}E_{n0}}+\frac{1}{E_{n0}^{2}E_{m0}}\right\} \right),\label{eq:gamma}
\end{equation}
\end{widetext}which contains three kinds of quantity: the energy
level differences $E_{nm}=E_{n}-E_{m}$, matrix elements $x_{nm}=\left\langle n\left|x\right|m\right\rangle $
and barred quantities $\bar{x}_{nm}$ that contain dipole terms,
\begin{equation}
\bar{x}_{nm}=\begin{cases}
\Delta x_{n0}=x_{nn}-x_{00} & ,\ n=m\\
x_{nm} & ,\ n\neq m
\end{cases}.\label{eq:dipole}
\end{equation}
As is well known, the dipole terms can be eliminated from these expressions
using the sum rules\cite{perezmoreno08,kuzyk05}, leaving only the
transition elements $x_{nm}$ and energy level differences $E_{nm}$.
Numerous previous studies, as well as the results above, have shown
that only a few states\textemdash in fact 2-3\textemdash near the
Fermi energy contribute significantly to the hyperpolarizabilities.
Hence, the question of what is important to the hyperpolarizability
maybe be addressed by understanding how the potential and wavefunctions
affect these quantities. In the remainder of the appendix, we will
answer this question by constructing WKB ansatz wavefunctions, inserting
them into expressions for these quantities and examining the form
of the results. 

In order to proceed, we make some simplifying assumptions. First,
we assume that the true optimum potential is sufficiently smooth that
the WKB approximation yields good approximations, at least near the
turning points, to the relevant wavefunctions, i.e. those near the
Fermi surface. This is justified by previous studies\cite{Wiggers07}
that have shown the addition of small rapidly varying perturbations
to the potential doesn't affect the hyperpolarizabilities. The ansatz
potentials studied in this work, with delta functions and changes
in slope at isolated points, all satisfy this criterion. 

Second, we shall assume that for wavefunctions near the Fermi energy
there are only two classical turning points, except possibly for isolated
delta functions in the potential that may violate this rule. We justify
this because the presence of multiple turning points would result,
at least approximately, in roughly independent particles in the separate
classically allowed regions. This would lead to hyperpolarizabilities
that grow $\propto N$ rather than $\propto N^{\frac{3}{2}}$ (for
$\beta$) or $\propto N^{2}$ (for $\gamma$) as the Kuzyk bounds
imply. The turning points shall be denoted $x_{n}^{l}$ and $x_{n}^{r}$,
corresponding to the left and right turning point respectively, and
are found by solving $V(x_{n}^{l})=E_{n}$ . In this expression, and
hereafter, $l$ indexes the left or right turning point.

Within the above assumptions it is possible to write an ansatz wavefunction
for the $n$th state that is valid except for the region near the
classical turning points, 
\begin{equation}
\psi_{n}(x)=\begin{cases}
A_{n}(x)\cos\left[\phi_{n}(x)-\Delta\phi_{n}(x)\right], & E>V\\
A_{n}(x)\exp\left[-\gamma_{n}^{l}(x)\right], & E<V
\end{cases}\label{eq:psi}
\end{equation}
where $E>V$ and $E<V$ refer to the classically allowed and forbidden
regions respectively. The functions $\gamma_{n}^{l}$ are defined
by,
\begin{equation}
\gamma_{n}^{l}(x)=\left|\int_{x_{n}^{l}}^{x}\left(V(x')-E_{n}\right)^{\frac{1}{2}}dx'\right|,\label{eq:WKBgamma}
\end{equation}
and the function $\phi_{n}(x)$ is given by,
\begin{equation}
\phi_{n}(x)=\frac{\pi}{4}+\int_{x_{n}^{<}}^{x}\left(E_{n}-V(x')\right)^{\frac{1}{2}}dx'.\label{eq:WKBphi}
\end{equation}
The remaining functions $A_{n}(x)$ and $\Delta\phi_{n}(x)$ are smooth
and slowly varying, except where the potential has delta functions
or sharp changes; nonetheless they are necessarily smoother than the
potential. For the smooth potentials considered here, $\Delta\phi(x)$
is of order $\pi$ or smaller. $A(x)$ is relatively close to the
semiclassical result $A(x)\approx a(x)v(x)^{\frac{1}{2}}$ where $v(x)$
is the classical velocity and $a(x)$ changes only by amounts that
are asymptotically small where the WKB approximation is valid, except
very close to the turning points. The dependence of $a(x)$ can easily
be made more precise by using the uniform asymptotic WKB approximation
to the wavefunction\cite{bender1999} though this does little to illuminate
the discussion. Moreover, even if WKB is invalid somewhere between
the turning points, the magnitude, $|a(x)|$ is, roughly speaking,
the rate at which the electron \textquotedbl{}turns\textquotedbl{}
at the turning point and so is expected to be the same at both turning
points. In WKB there is no reflection in classically allowed regions
and hence it is obvious that $|a(x)|$ cannot change. More generally,
however, any structure between the turning points must have equal
magnitude for the transmission and reflection from each side. This
implies that $|a(x)|$ must have the same values at both turning points.
Moreover, $a(x)$ can be chosen to be real near both turning points,
and the ratio of $a$ near the two turning points alternates sign
as we increase the energy of the wavefunctions.

\begin{widetext}

The approximate energies are found, as usual, by solving

\begin{equation}
n+\frac{1}{2}=\phi(x_{n}^{>})=\int_{x_{n^{l}}}^{x_{n}^{r}}\left(E_{n}-V(x')\right)^{\frac{1}{2}}dx'.\label{eq:WKBequation}
\end{equation}

Combining copies of the WKB equation (\ref{eq:WKBequation}) for two
different states $n$ and $m$, we obtain,

\begin{equation}
n-m=\int_{x_{n^{l}}}^{x_{n}^{r}}\left(E_{n}-V(x')\right)^{\frac{1}{2}}dx'-\int_{x_{m}^{l}}^{x_{m}^{r}}\left(E_{m}-V(x')\right)^{\frac{1}{2}}dx'.\label{eq:WKBnm}
\end{equation}
These integrals may be combined by making a linear change of variable
$x'\to x_{p}^{l}+y\Delta x_{p}$ where $y\in[0,1]$ and $\Delta x_{p}=x_{p}^{r}-x_{p}^{l}$.

\begin{equation}
n-m=\int_{0}^{1}dy\left[\Delta x_{p}\left[E_{n}-V\left(x_{n}^{l}+y\Delta x_{p}\right)\right]^{\frac{1}{2}}-\Delta x_{p}\left[E_{m}-V\left(x_{m}^{l}+y\right)\right]^{\frac{1}{2}}\right].\label{eq:WKBnm2}
\end{equation}
The square roots in the integrand can be combined by completing the
square, 

\begin{equation}
n-m=\int_{0}^{1}dy\frac{\Delta x_{n}^{2}\left[E_{n}-V(\Delta x_{n}y+x_{n}^{l})\right]-\Delta x_{t,m}^{2}\left[E_{m}-V(\Delta x_{m}y+x_{m}^{l}\right]}{\Delta x_{n}\left[E_{n}-V(\Delta x_{n}y+x_{n}^{l})\right]+\Delta x_{t,m}\left[E_{m}-V(\Delta x_{m}y+x_{m}^{l})\right]}.\label{eq:WKBnmFINAL}
\end{equation}

The form of the integrand in (\ref{eq:WKBnmFINAL}) is instructive:
near the turning points, i.e. as $y\to0$ and $y\to1$, the numerator
goes to zero linearly in $y$ while the denominator goes to zero as
a square root; the integrand therefore vanishes like $y^{1/2}$ near
the turning points. Hence, the majority contribution to this integral
comes from the spatial region far from the classical turning points,
particularly if the two energies are similar and the potential near
the turning points is slowly varying. Moreover, this integral tends
to smooth out small, high frequency variations in the potential; it
follows that if $\left|n-m\right|$ is small, then $E_{n}-E_{m}$
is largely determined by the form of the potential far from the turning
points. 

We now turn to the dipole matrix elements $x_{nm}$, which can be
computed using the position formula,
\begin{equation}
x_{nm}=\int_{-\infty}^{\infty}dxx\psi_{n}\psi_{m}.\label{eq:POSITION}
\end{equation}
We shall restrict our analysis to small values of $\left|n-m\right|$,
because only these states contribute significantly to the hyperpolarizability.
Moreover, approximate analysis of these integrals is complicated for
large $\left|n-m\right|$ because, while small, they are strongly
dependent on the detailed analytic behavior of the potential and wavefunctions.

Substituting the ansatz wavefunction (\ref{eq:psi}) into the position
formula (\ref{eq:POSITION}), 
\begin{equation}
x_{nm}\approx\int_{x_{m}^{<}}^{x_{m}^{>}}dx\frac{1}{2}A_{n}A_{m}\left[x-\frac{1}{2}(x_{t,m}^{r}+x_{t,n}^{l})\right]\cos(\phi_{n}-\phi_{m}+\Delta\phi_{m}-\Delta\phi_{m}),\label{eq:positionintegral}
\end{equation}
\end{widetext}. This integral is then the integral of a function
that has $n-m-1$ nodes, roughly evenly spaced through the interval,
and has a somewhat (algebraically) larger magnitude near the turning
points. From this formula, it is apparent that the transition matrix
elements depend primarily on the nature of the wavefunctions near
the classical turning points, i.e. separation between the two turning
points and as well as the shape of the wavefunctions in their vicinity.
As the shape of the wavefunctions are most dependent on the slopes
of the potential at the turning points and there must be a connection
between the energy differences and the transition matrix elements,
this strongly argues that the separation between the turning points
of the HOMO and the slopes of the potential at these turning points
should be among the most important heuristic parameters that determine
the hyperpolarizabilities.

\end{document}